\documentclass[aps,prb,reprint,superscriptaddress,showpacs]{revtex4-1}

\usepackage{graphicx}
\usepackage{amssymb}
\usepackage{amsmath}
\DeclareGraphicsExtensions{.eps}

\begin{document}

\title{Contact-free probing of interfacial charging and Debye-like charge screening in SiO$_2$/PDI8-CN$_2$ systems by optical second harmonic generation}

\author{F. Ciccullo}
\affiliation{Physics Department, University of Naples “Federico II”, Via Cintia I-80126 Napoli (Italy)}
\affiliation{Institute for Superconductors, Oxides and Innovative Materials, National Research Council (CNR-SPIN), U.O.S. Napoli, Via Cintia, I-80126 Napoli (Italy)}

\author{A. Cassinese}
\affiliation{Physics Department, University of Naples “Federico II”, Via Cintia I-80126 Napoli (Italy)}
\affiliation{Institute for Superconductors, Oxides and Innovative Materials, National Research Council (CNR-SPIN), U.O.S. Napoli, Via Cintia, I-80126 Napoli (Italy)}

\author{E. Orabona}
\affiliation{Physics Department, University of Naples “Federico II”, Via Cintia I-80126 Napoli (Italy)}
\affiliation{Institute for Superconductors, Oxides and Innovative Materials, National Research Council (CNR-SPIN), U.O.S. Napoli, Via Cintia, I-80126 Napoli (Italy)}

\author{L. Santamaria}
\thanks{Present address: INO-CNR, Istituto Nazionale di Ottica, Sezione di Napoli, Via Campi Flegrei 34, I-80078 Pozzuoli (NA); Italy}
\affiliation{Physics Department, University of Naples “Federico II”, Via Cintia I-80126 Napoli (Italy)}

\author{P. Maddalena}
\affiliation{Physics Department, University of Naples “Federico II”, Via Cintia I-80126 Napoli (Italy)}
\affiliation{Institute for Superconductors, Oxides and Innovative Materials, National Research Council (CNR-SPIN), U.O.S. Napoli, Via Cintia, I-80126 Napoli (Italy)}

\author{S. Lettieri}
\email[To whom correspondence should be addressed. E-mail: ]{stefano.lettieri@spin.cnr.it}
\affiliation{Institute for Superconductors, Oxides and Innovative Materials, National Research Council (CNR-SPIN), U.O.S. Napoli, Via Cintia, I-80126 Napoli (Italy)}



\begin{abstract}
Investigation of the interfacial electronic properties of N,N'-bis(n-octyl)-(1,7\&1,6)-dicyanoperylene-3,4:9,10-bisdicarboximide (PDI8-CN$_2$) organic semiconductor films grown on silicon dioxide is performed by polarization-resolved second harmonic generation optical spectroscopy, pointing out a spatial region where charge carriers distribution in the semiconductor lacks inversion symmetry. By developing a model for nonlinear susceptibility in the framework of Debye-Huckel screening theory, we show that the experimental findings can be interpreted as resulting from the presence of a net charge localized at the silicon dioxide, accompanied by a non-uniform charge distribution in the organic semiconductor.  Photoluminescence analysis further reinforces this scenario. Reduction-oxidation reactions involving PDI8-CN$_2$ and water molecules are invoked as physical origin of the localized charge. The work outlines a sensitive tool to probe the total charge localized at buried semiconductor/dielectric interfaces in organic thin-film transistors without resorting to invasive contact-based analyses.
\end{abstract}

\maketitle

\section{Introduction}
\label{}

It is commonly recognized that realization of full potential of organic electronics is closely related on the ability to accurately analyze and predict the electronic properties of the various interfaces that play a role in the devices.\cite{Ishii99,Braun09} Such concept is particularly evident in the case of organic field-effect transistors (OFETs), whose performances are decisively affected by local interactions of the organic semiconductor at its interface with gate dielectric\cite{Veres03,Stassen04,Hulea06,Sirringhaus09,Zilker01,Richards08,Sharma10} and air.\cite{Chen12,DeLeeuw97,Newman04}.

The important phenomenon of voltage threshold shift during continuous gate bias (also known as “bias stress effect”) can be considered a representative example of the influence exerted on OFETs charge transport properties by the organic semiconductor/gate dielectric interplay.\cite{Kobayashi04} Bias stress effect is commonly attributed to charge trapping occurring in the semiconductor/dielectric transitional region and the issue of actual location (either semiconductor or gate dielectric) of the fixed charges has remained elusive for some time,\cite{Mathijssen07} until the possibility to probe the trapped charge in a direct manner has been demonstrated by semiconductor removal through scotch-tape exfoliation followed by scanning Kelvin probe microscopy analysis of the exposed dielectric layer.\cite{Mathijssen07,Andringa12} Even if these analysis are very convincing and important, they involve some technical complications (e.g. films need to be mechanically robust to allow a complete exfoliation, ambient light can de-trap the charge) and, in particular, the analysis implies the device destruction. Ultimately, such complications can be avoided only by developing a non-contact and spatially-selective probing of the buried interface.

Optical analysis can indeed provide a direct and contact-free probing of electronic processes,\cite{DiPietro12a,DiPietro12b} but it is also to be noted that many of the most common optical spectroscopy techniques are based on linear matter-radiation interaction processes (such as absorption, reflection, fluorescence, etc.) taking place in the whole bulk of the medium: as a consequence, they are not intrinsically spatially-selective. Exploitation of optical processes carrying an intrinsic sensitivity to the physical state of surfaces and/or interfaces is instead desirable: one example of such processes is represented by optical second harmonic generation (SHG), consisting in the generation of an electromagnetic wave of frequency 2$\omega$ (“second-harmonic” or SH wave) caused by the interaction between a material medium and a laser beam (“fundamental beam”) of frequency $\omega$.
Remarkably, symmetry considerations allow to demonstrate that such process cannot take place in material regions where inversion symmetry holds.\cite{Butcher91} Therefore, SHG analysis is intrinsically spatially-selective, as it allows probing electronic properties related only to regions where the inversion symmetry is broken by material discontinuities (surfaces, interfaces) or by the presence of local field having a well-defined direction.

As being an optical spectroscopy technique, SHG analysis allows to exploit peculiarities such as access to interfaces, absence of physical contacts, possibility to study dynamic processes by time-resolved detection.\cite{Lettieri05,Lettieri07,Chang97,Guo01,Hoffmann11} Also thanks to such peculiarities, SHG spectroscopy has been successfully employed in last years in the field of organic field-effect transistors for investigations on carrier mobility,\cite{Manaka07,Manaka08} channel formation,\cite{Lim08} trap states\cite{Tanaka11} and charge transfer at heterostructures.\cite{DiGirolamo12}

Driven by such considerations, we report about the first SHG analysis - assisted by photoluminescence (PL) investigations - devoted to the study of interfacial charge trapping and internal built-in electric field formation in N,N'-bis(n-octyl)-(1,7\&1,6)-dicyanoperylene-3,4:9,10- bisdicarboximide (PDI8-CN$_2$) films grown on silicon dioxide (SiO$_2$). PDI8-CN$_2$ is a perylene diimide derivative considered to be one of the most promising materials for realization of stable n-channel OFETs\cite{Jones04,Jones08} and complementary circuits,\cite{Yoo06} also yielding high-performance devices by sublimation or from solution methods.\cite{Rivnay09} Recent investigations allowed to establish that the molecular packing in PDI8-CN$_2$ films is characterized by planar centrosymmetric perylene cores stacked cofacially, with π–π interactions almost perpendicular to the substrate surface.\cite{Liscio12} Thus, SHG spectroscopy appears to be particularly suited to the study of surface/interface processes in PDI8-CN$_2$ considering its centrosymmetry and its in-plane charge delocalization.\cite{Flytzanis95}

Analyzing a set of variable thickness PDI8-CN$_2$ thin films by polarization-resolved SHG (PR-SHG) and extracting the amplitude of $\chi_{zxx}$ second-order dielectric susceptibility tensor elements, a finite three-dimensional PDI8-CN$_2$ region where the spatial distribution of charge carriers lacks inversion symmetry has been evidenced. A model to interpret this finding has been developed on the basis of a Debye-Huckel charge screening mechanism triggered by charges localized in the SiO$_2$ dielectric layer. Such picture, confirmed by photoluminescence findings, is used here to develop an expression of electric field-induced SHG (EFISHG) susceptibility associated to the screening field, allowing extracting an effective Debye length for the charge distribution induced in the semiconductor by the fixed charge located in the gate dielectric.

The work is organized as follows: the experimental methods and details are described in Section 2;  Section 3 is devoted to the experimental results, involving PR-SHG and excitation-resolved photoluminescence (PLE) analysis; in Section 4 the interpretation model developed to elucidate the experimental results is described, followed by a discussion on PL-PLE results and on other literature findings. Main conclusions are finally drawn in section 5.

\section{Experimental methods and details}
Experimental investigation in the present work was carried out on a set of five different thickness PDI8-CN$_2$ thin films deposited on SiO$_2$, indicated hereafter as S1, S2, S3, S4, S5 and having a thickness ($d$) of 6 nm, 10 nm, 20 nm, 42 nm and 70 nm respectively. The PDI8-CN$_2$ powder was purchased from Polyera Corporation Inc. (Polyera ActivInkTM N1200) and deposited by vacuum sublimation on 200 nm-thick SiO$_2$ layers thermally grown on commercial (001) silicon substrates. Depositions were performed at a growth rate of 6 $\AA$ min$^{-1}$ by a high-vacuum system (base pressure between $10^{-8}$ and $10^{-7}$ mbar) equipped with a Knudsen cell and a quartz microbalance. The chamber was warmed at temperature T=$90^{\circ}$ during deposition. Post-growth morphological characterization of films was performed by means of a XE100 Park atomic force microscope operating in air.

Prior to PR-SHG analysis, PL and PLE spectroscopy measurements were performed on samples in order to gain information on optical transition levels and on presence of near band-edge states in PDI8-CN$_2$ films: as discussed in section 4, such information is helpful to interpret the SHG experimental results. The PL and PLE analysis was performed by using a computer-controlled motorized system composed of a broad-band emission Xe lamp coupled with a monochromator equipped with 1200 grooves/mm double-grating, providing wavelength-tunable monochromatic optical excitation. PL emission was collected through a confocal lens system and focused onto the input slit of a 320 mm focal length spectrometer. The PLE signal was determined as the total PL emission (i.e. photoluminescence intensity spectrum integrated over emission photon energy) obtained as a function of excitation photon energy.

\begin{figure}
\includegraphics{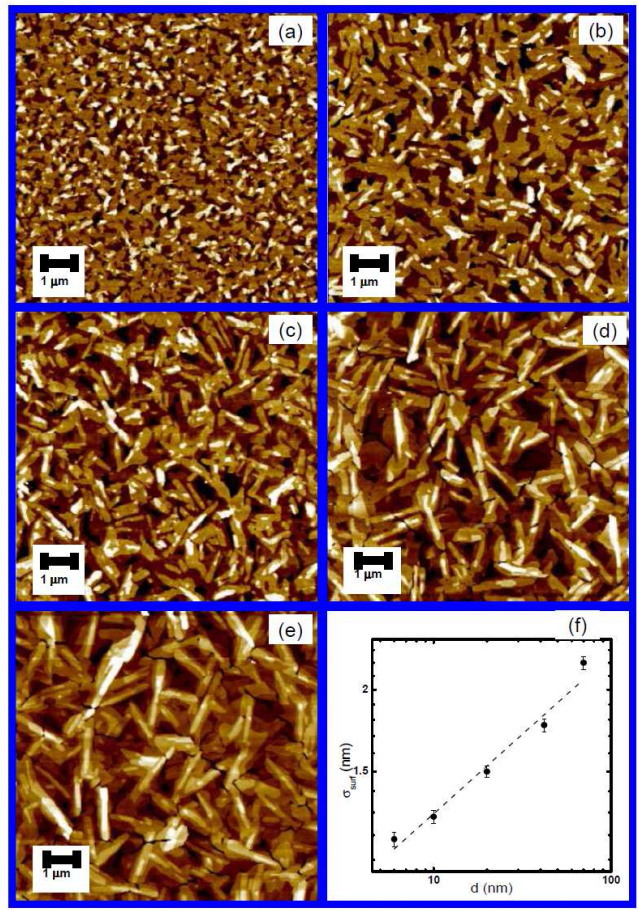}
\caption{Topographic AFM images of PDI8-CN2 films of different thickness: (a): 6 nm (sample S1); (b): 10 nm (sample S2); (c): 20 nm (sample S3); (d): 42 nm (sample S4); (e): 70 nm (sample S5). In panel (f) the surface roughness is reported vs. film thickness in log-log scale. }
\end{figure}

PR-SHG analysis was performed in reflection geometry (angle of incidence of about $45^{\circ}$°) by using a Nd:YAG mode-locked laser (emission wavelength $\lambda_{\omega}$=1064 nm, 20 ps pulses duration, 10 Hz repetition rate) as fundamental beam. The polarization angle $\alpha$ of the linearly-polarized laser beam was varied through a 1064 nm half-wave retardation plate mounted on computer-controlled motorized rotation stage. Sharp cut-off optical filters were used to block fundamental beam reflected by the sample and the undesired residual SH wave generated in the various optical elements composing the setup. The p-polarized component of SH output beam was selected by rotating a half-wave retardation plate operating at SH wavelength ($\lambda_{2\omega}$=532 nm) and placed before of a fixed polarizer. Finally, the output radiation was spectrally filtered at SH wavelength by a PC-controlled motorized monochromator in order to remove any eventual spurious signal. A photomultiplier tube was used for SHG intensity detection. Each experimental point was obtained by averaging the SHG signal over 400 laser shots.

\section{Experimental results}

Topographic AFM images of as-deposited samples acquired on 10x10 $\mu m^2$ area are shown in Fig. 1 (panels (a)-(e)) evidencing an uniform quasi-two-dimensional growth for thinnest films, followed by progressive formation of elongated three-dimensional islands about one micrometer long at increasing film thickness, in accord with other literature findings.\cite{Liscio12,Chiarella11} By computing the root-mean-square roughness σsurf (determined as the standard deviation of the film height distribution) versus film thickness d, a power law $\sigma_{surf}\propto d^\beta$ is obtained, in accord to dynamic scaling theory.\cite{Krug04} A best-fit value of $\beta=0.24\pm0.03$ was obtained for the growth exponent (Fig. 1, panel (f)), in close accordance with the one ($\beta=0.21\pm±0.02$) already reported by Liscio and coworkers for PDI8-CN$_2$ grown in similar conditions and representing a typical value for compact surfaces developing by means of a layer-by-layer growth mechanism.\cite{Liscio12}

\begin{figure}
	\includegraphics{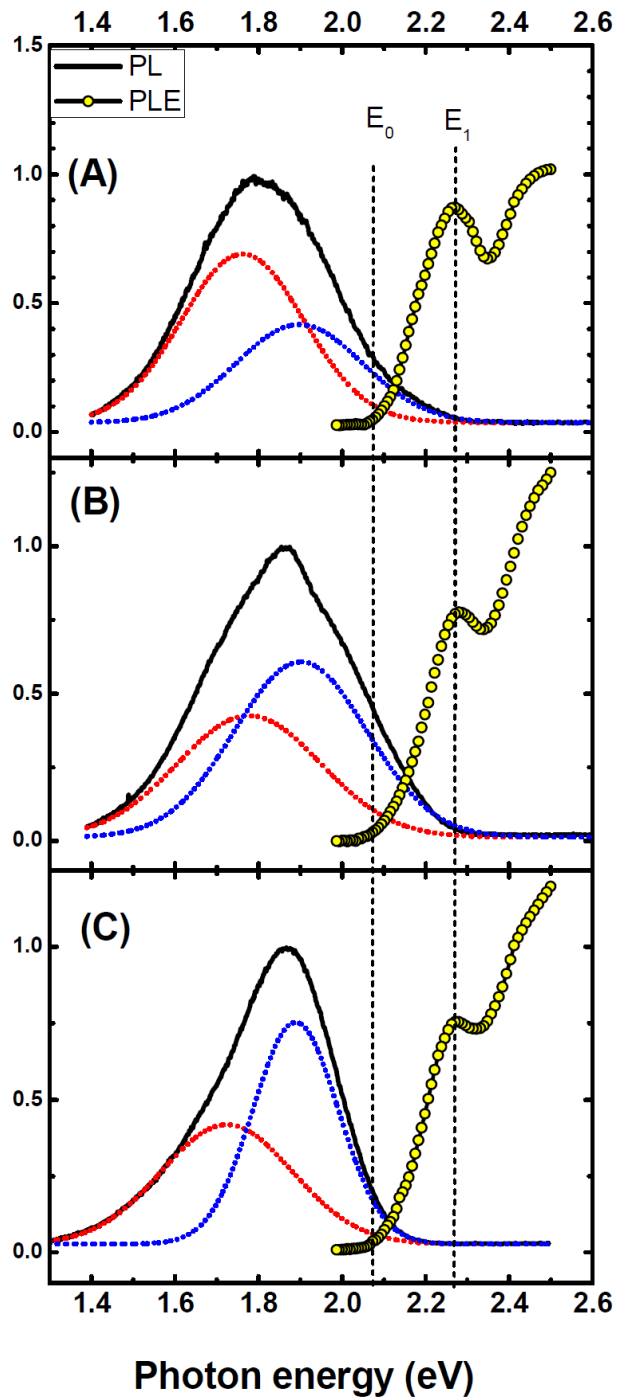}
	\caption{(Color online) Peak-normalized PL (black curves) and PLE (yellow circles, arbitrary units) spectra of samples S3 (panel A), S4 (panel B) and S5 (panel C). The dotted curves are the Gaussian component used to decompose the PL spectra, obtained by double-Gaussian best fit of the PL data. To help comparison, approximate energy position of the excitation onset ($E_0\approx$ 2.08 eV) and of the first excitation resonance ($E_1\approx$ 2.28 eV) are marked by vertical dashed lines.}
\end{figure}

PL and PLE characterizations were performed through the experimental setup previously described. The excitation photon energy in PLE analysis was varied from 1.98 eV (corresponding to 625 nm excitation wavelength) to 2.5 eV (corresponding to 496 nm excitation wavelength), thus scanning the energy range from sub-bandgap to interband (i.e. HOMO-LUMO) transition energies and allowing to determine the spectral shape of optical absorption of organic films.\cite{Chiarella11}
Similar spectral features in the PL emission spectra were observed in all samples. In particular, all PDI8-CN$_2$ films exhibited broad emission spectra approximately peaked at ≈ 1.85 eV photon energy and whose profiles were not symmetric with respect to their centre. Representative peak-normalized PL spectra are reported in Fig. 2 (solid black curves) for samples S3 (top panel, 2A), S4 (middle panel, 2B) and S5 (bottom panel, 2C). It can be recognized that the centre of mass of the emission profile does not coincide with the peak position, suggesting therefore the presence of more than a single PL band. The spectra could indeed be satisfactorily described as superposition of two Gaussian emission bands, shown in Fig. 2 as dotted curves. Finally, the excitation spectra obtained by PLE measurements are reported in Fig. 2 as open circles.
To help comparison between the different plots, vertical dotted lines have been inserted to mark the energy positions of the excitation edge ($E_0$) and of the first excitation resonance (corresponding to optical absorption peaks) indicated as $E_1$. It is worth underlining that no significant sample-to-sample variation in the latter values is evidenced. As discussed in next section, the observed features – namely: a) the occurrence of an asymmetric PL profile, with likely presence of double emission bands; b) the relevant spectral red-shift between the PL peak and the absorption edge, and; c) the partial overlap between high-energy tail of PL emission and absorption edge – will be useful to sketch a scheme for distribution of occupied electronic levels in PDI8-CN$_2$ films, supporting in turn the interpretation of SHG data.

\begin{figure}
	\includegraphics{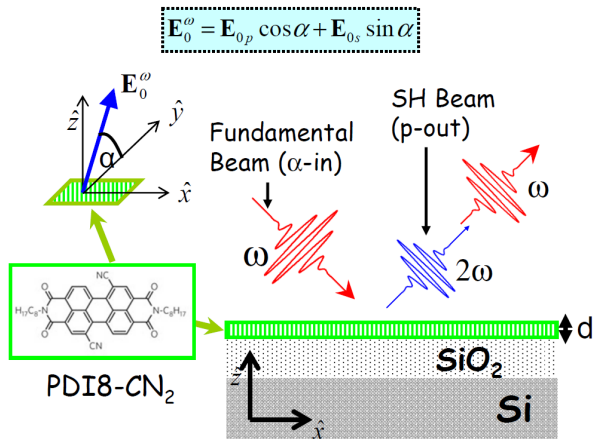}
	\caption{(Color online) Schematic representation of the SHG experiment geometry. The fundamental beam of frequency $\omega$ impinges on sample at an angle of incidence $\theta$. Reflected fundamental beam and SH beam are both represented in the figure, where in the actual experimental setup the former is cut off by a suitable optical filter. The sample surface and plane of incidence are parallel to $xy$ and $xz$ planes respectively. The polarization angle $\alpha$ is defined as the angle formed by the polarization direction of the fundamental electric field and the horizontal $xy$ plane. In the green frame the structural formula of PDI8-CN$_2$ molecule is reported.}
\end{figure}

The PR-SHG experiment geometry is sketched in Fig. 3, showing the polarization angle $\alpha$ (defined as the angle between the polarization direction of the linearly-polarized fundamental electric field $\textbf{E}_\omega$ and the sample surface) and the reference system used to describe the SHG signal. The experiments were performed by detecting the optical intensity of the reflected p-polarized (i.e. parallel to the plane of incidence $xz$ in Fig. 3) SH wave as a function of the polarization angle $\alpha$. The general purpose of the PR-SHG approach is to extract from experimental data the individual SHG susceptibilities (i.e. tensor elements of the $\chi^{(2)}$ tensor that drives the SHG process), as they represent the physical quantities that carry information on both surface properties and electric field-induced effects. In the case of isotropic media, $\chi^{(2)}$ tensor is composed by only three independent non-null elements, namely $\chi_{zzz}$, $\chi_{zxx}$ and $\chi_{xxz}$. Using the axis representation shown in Fig. 3, the polar pattern of p-polarized SHG intensity ${I_P} \propto {\left| {{E_x}} \right|^2} + {\left| {{E_z}} \right|^2}$ vs. $\alpha$ (indicated also as “$\alpha$p plot”) can be directly related to the non-null susceptibility elements as follows:\cite{Corn94}
\begin{equation}
{I_p^{2\omega }(\alpha ) \propto {\left( {{I_\omega }\sin \theta } \right)^2} \cdot {\left| {A{{\sin }^2}\alpha  + B} \right|^2}}
\label{eq:1}
\end{equation}
where $B=\chi_{zxx}$: therefore, the $\chi_{zxx}$ coefficient can be singled out through a best-fit of $\alpha$p plots, while the term A has a more complicated expression, combining all the susceptibility elements with trigonometric function of the angle of incidence.\cite{Guyot86,Zhang90}

In Fig. 4 (upper panel) the experimental $\alpha$p plots are reported for some of the investigated samples (to improve the readability of the graph the $\alpha$p curve of the S3 sample is omitted). All of the investigated PDI8-CN$_2$ films indeed exhibited the same $\alpha$p polar pattern, in accordance with Eq.\ref{eq:1} and differing only in actual values of the SHG intensity (and, thus, in actual values of the best-fit coefficients). In other words, all investigated PDI8-CN$_2$ films exhibited an SHG polar pattern representative of an average isotropic symmetry extended over the size of the optical interaction spot. Observation of such planar isotropy was indeed predictable, after considering that the optical excitation probes a surface area including a very high number of elongated domains with random in-plane orientation. As a rough estimation, we can take the in-plane correlation length $\xi$ as a characteristic average size of an ordered domain: considering typical $\xi$ values ranging in PDI8-CN$_2$ from about 80 to 150 nm,\cite{Chiarella11} an estimated number of 10$^6$ randomly-oriented domains are probed on average in our measurements, thus smoothing out local in-plane anisotropy of individual domains.

\begin{figure}
\includegraphics{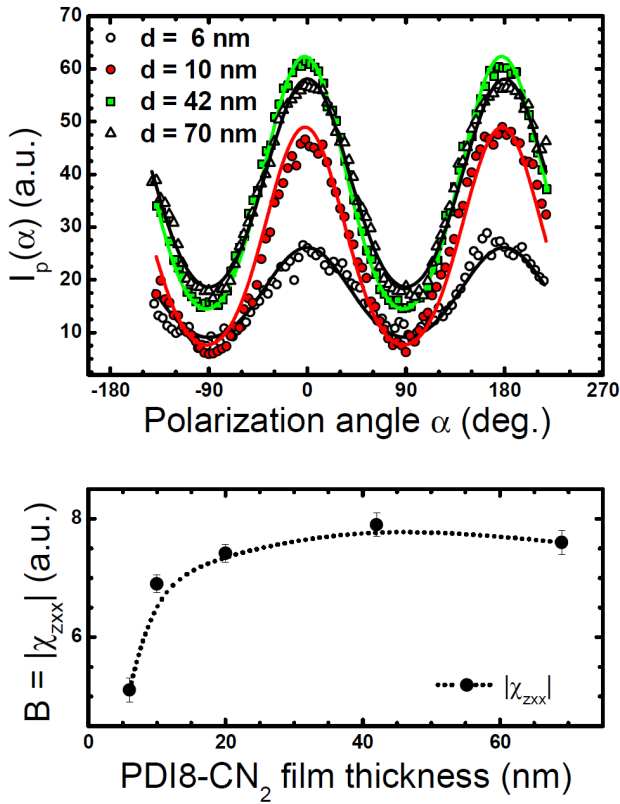}
\caption{(Color online) Top panel: $\alpha$-in/p-out SHG intensity for samples S1, S2, S4 and S5. Full lines represent the best-fit curves obtained by using Eq. (1). Bottom panel: $\left|\chi_{zxx}\right|$ susceptibility obtained by best fit of PR-SHG $\alpha$-in/p-out data using Eq.(1) vs. PDI8-CN$_2$ film thickness. The dotted line is a guide to the eye.}
\end{figure}

Best-fit curves obtained by using Eq. (1) are reported in Fig. 4 (upper panel) as solid black curves. The corresponding best-fitting values for $\left|B\right|=\left|\chi_{zxx}\right|$susceptibility coefficient are plotted in the lower panel of Fig. 4 as a function of PDI8-CN$_2$ films thickness. It can be seen that the SHG susceptibilities exhibit a peculiar behavior, characterized by a thickness dependence which is marked for smaller film thicknesses and fades into a plateau at larger film thicknesses. Even at this pre-analysis stage, it can thus be anticipated that the results strongly suggest that the observed nonlinear response cannot be assigned to interface-localized  processes only, which are not expected to depend on film thickness.

\section{Discussion}
\subsection{Interpretation of experimental results: simplified Debye-Huckel model for SHG susceptibility in presence of an interfacial charge layer}

As previously mentioned, the observed dependence of PDI8-CN$_2$ SHG intensities and susceptibilities on sample thickness suggests that interfacial-related optical nonlinearity is not the sole contribution to the second harmonic signal. Indeed, the behavior represented in the bottom panel of Fig 4 indicates that the “SHG-active” region is not limited to only one or two molecular layers as in “standard” surface/interface-related SHG, but indeed extends for several nanometers,  defining a finite three-dimensional volume corresponding to a semiconductor thickness above which the SHG signal saturates.
Actually, inversion symmetry in centrosymmetric semiconductors can be broken not only by material discontinuities - leading to the properly called surface/interface SHG - but also by local fields having a well-defined direction, leading in this case to the electric field-induced SHG (EFISHG) phenomenon. It is worth underlining that the two situations involve different SHG-active regions, namely a quasi-two dimensional transitional region in the former case and a three-dimensional region (in which local electrostatic fields are present) in EFISHG.

In this section we show that the observed features can be actually interpreted in terms of EFISHG induced by a non-uniform charge distribution related to the presence of a charged interfacial layer.
To this aim, we start by developing the nonlinear susceptibility term associated to electric field-induced SHG phenomenon. The latter is a third-order phenomenon, driven by the $\boldsymbol{\chi}^{(3)}$ susceptibility tensor:\cite{Lee67} as being an odd-order term, $\boldsymbol{\chi}^{(3)}$ is not restricted by inversion symmetry considerations and is therefore non-vanishing in the bulk of centrosymmetric media. The EFISHG term arises from the local presence of a static electric field (indicated by $\mathbf{E}^{dc}(z)$) giving rise to a nonlinear polarization whose expression (using tensor notation) is \cite{Shen02,Aktsipetrov94}
$\mathbf{P}_{2\omega }^{(EFISHG)}=\boldsymbol{\chi}^{(3)}:\mathbf{E}^\omega\mathbf{E}^\omega\mathbf{z}\cdot E^{dc}(z)$, where $\mathbf{E}^{dc}$ is assumed to be directed along the normal to material surface due to the isotropy of the system in the (xy) plane. The previous expression clearly indicates that the EFISHG polarization represents an additional SHG contribution, originating from the presence of a $E^{dc}(z)$ term that breaks the symmetry under inversion of the z-axis.

For the present work, we derived the expression of the total effective SHG susceptibility in presence of both interfacial and electric field-induced SHG by using the Green function formalism developed by Sipe for surface optics.\cite{Sipe87} The details for the calculation of the effective $\chi_{zxx}$ susceptibility (which is the quantity determined though the experimental $\alpha$p polar plots) for the case of an isotropic film are reported in Appendix, where the following expression is demonstrated:
\begin{equation}
{\chi _{zxx}} = \chi _{zxx}^{(I)} + {\gamma _3}\exp \left( { - {d \mathord{\left/
 {\vphantom {d \Lambda }} \right.
 \kern-\nulldelimiterspace} \Lambda }} \right) \cdot F(d)
\label{eq:2}
\end{equation}
where $d$ is the semiconductor thickness, $\chi_{zxx}^{(I)}$  represents the interfacial second-order nonlinear susceptibility and $\gamma_3$ is the $zxxz$ element of $\chi^{(3)}$ tensor element (i.e. one of the three non-null independent tensor elements of ${{\boldsymbol{\chi }}^{(3)}}$ tensor characterizing an isotropic medium, see Appendix for the exact definition). The actual spatial profile of the internal electrostatic field and the semiconductor thickness affect the effective SHG susceptivity through the term $F(d)$, whose expression is:

\begin{equation}
{F(d) = \int_0^d {{E^{dc}}(z)\exp \left[ { - {{iz} \mathord{\left/
 {\vphantom {{iz} {{L_C}}}} \right.
 \kern-\nulldelimiterspace} {{L_C}}}} \right]dz}}
\label{eq:3}
\end{equation}
where $L_C=\lambda_{\omega}/{4\pi \left( n_{\omega} \cos\theta_{\omega} + n_{2\omega}\cos \theta_{2\omega} \right)}$, $\theta_{i}$ ($i=\omega, 2\omega$) is the refraction angle of fundamental ($i=\omega$) and SH wave ($i=2\omega$) in the sample, $n_{i}$ is the PDI8-CN$_2$ refractive index at fundamental wavelength ($\lambda_{\omega}$) and SH wavelength $\lambda_{2\omega}$. Finally, 
$\Lambda =\lambda_{2\omega}/{2\pi \kappa_{2\omega}\cos\theta_{\omega}}$,where $\kappa_{2\omega}$ is the extinction coefficient of PDI8-CN$_2$ at SH wavelength. 

On the basis of the above expression, we show here that the peculiar behaviour obtained for SHG susceptibility $\left|\chi _{zxx}\right|$ in SiO$_2$/PDI8-CN$_2$ systems can be interpreted as resulting from the presence of a net charge in SiO$_2$ gate dielectric, localized about its interface with the organic semiconductor. To this aim, it is to be underlined that the organic molecules are in the neutral state in SHG experiments: as a consequence, accumulation of localized charges has to be compensated by redistribution of opposite charge, screening out the local charge and guaranteeing the overall charge neutrality of the organic semiconductor. The built-in electrostatic field ${E^{dc}}$ induced by the redistributed charge can thus give rise to an EFISHG term in the total SHG response, according to Eq.(\ref{eq:2}) and Eq.(\ref{eq:3}).

The spatial profile ${E^{dc}}(z)$ of the electrostatic field  is of course ruled by the Poisson equation
${{d{E^{dc}}} \mathord{\left/
 {\vphantom {{d{E^{dc}}} {dz =  - {{{d^2}\phi } \mathord{\left/
 {\vphantom {{{d^2}\phi } {d{z^2} = 4\pi \rho (z)}}} \right.
 \kern-\nulldelimiterspace} {d{z^2} = 4\pi \rho (z)}}}}} \right.
 \kern-\nulldelimiterspace} {dz =  - {{{d^2}\phi } \mathord{\left/
 {\vphantom {{{d^2}\phi } {d{z^2} = 4\pi \rho (z)}}} \right.
 \kern-\nulldelimiterspace} {d{z^2} = 4\pi \rho (z)}}}}$, where $\rho(z)$ indicates the density of redistributed charge in the organic semiconductor. In the general case, the latter has to be solved numerically\cite{Mayergoyz86,Pacelli97} due to its intrinsic non-linear nature (i.e. the charge density itself depends on the local value of electrostatic potential $\phi$). However, a simplified analytical expression for ${E^{dc}}(z)$ can be obtained under assumption of Debye-Huckel approximation, consisting in a linearization of the $\rho$ vs. $\phi$ dependence. In the case of a non-degenerated n-type semiconductor at temperature T, such approximation leads to a exponentially-decaying built-in electric field characterized by a screening length $L_D$:\cite{Luth10}

\begin{equation}
{E^{dc}}(z) = E_0^{dc}\exp \left( { - {z \mathord{\left/
 {\vphantom {z {{L_D}}}} \right.
 \kern-\nulldelimiterspace} {{L_D}}}} \right)
\label{eq:4}
\end{equation}
where ${L_D} = \sqrt {{{\varepsilon {k_B}T} \mathord{\left/ {\vphantom {{\varepsilon {k_B}T} {8\pi {e^2}{n_b}}}} \right. \kern-\nulldelimiterspace}{8\pi {e^2}{n_b}}}}$ is the Debye screening length, n$_b$ is the bulk density of mobile charge carriers and k$_B$ is the Boltzmann constant.

In spite of its simplicity, the above expression for the electric field takes into account the main features observed for SHG susceptibilities of PDI8-CN$_2$. In fact, using Eqs.(\ref{eq:2}), (\ref{eq:3}) and (\ref{eq:4}) the field-dependent quantity $F(d)$ becomes:

\begin{equation}
{F(d)=\int_{0}^{d}E_{0}^{dc}\exp \left[ -z\left( \frac{1}{L_{D}}+\frac{i}{L_{C}}\right) \right] dz}
\label{eq:5}
\end{equation}

It is easily seen that the modulus of Eq.(\ref{eq:5}) describes a thickness-dependent quantity that increases for $d<L_D$, followed by saturation and attenuation as film thickness approaches the attenuation length $\Gamma$, thus reproducing the behavior of experimentally determined quantity.

In the case where the built-in electric field extends up to a Debye length $L_D$ much less than the SHG coherence length $L_C$, a first-order Taylor approximation ($\exp \left[ -iz/L_{C}\right] \cong 1$) leads to the following simple expression:
\begin{equation}
\chi _{zxx}(d)=\chi _{zxx}^{(I)}+\gamma _{3}L_{D}E_{0}^{dc}\exp (-d/\Lambda )\left[ 1-\exp (-d/L_{D})\right]
\label{eq:6}
\end{equation}
while the imaginary part of $F(d)$ has to be considered for the $ \left|\chi_{zxx}(d)\right|$ calculation in the more general case of a Debye length not negligible compared to the coherence length $L_C$.

\begin{figure}
	\includegraphics{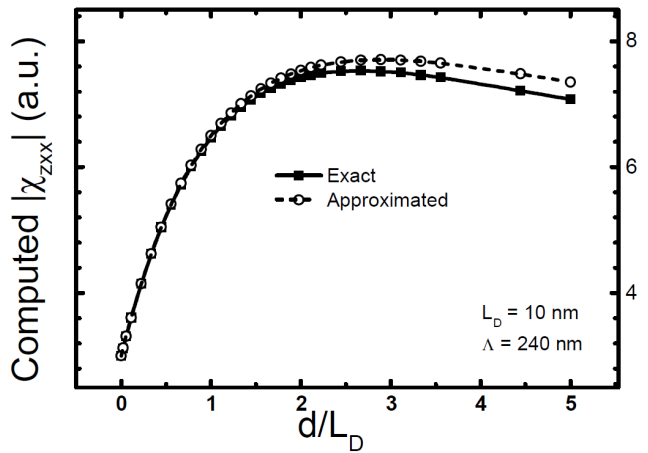}
	\caption{(Color online) Computed $\left|\chi_{zxx}\right|$ values obtained by using the approximated expression in Eq.(\ref{eq:6}) (empty circles) and the exact expression (full squares) as a function of $d/L_D$ ratio. The following numerical parameters have been used: $L_D = 10$ nm, $L_C = 30$ nm, $\Lambda = 240$ nm.}	
\end{figure}

We numerically computed the $\left|\chi _{zxx}(d)\right|$ values obtained through the exact (i.e. using Eq.(5)) vs. simplified (Eq.(6)) expression. The results are reported in Fig. 5, using the following parameter values: $L_D$ = 10 nm, $\Lambda$ = 240 nm, $L_C = 3\cdot L_D$ = 30 nm. The value for $L_D$ has been chosen as being close to the one obtained from the fit of experimental results (see next), while the values for $\Lambda$ and $L_C$ obtained using the ellipsometric values\cite{Tkachenko12} $n_\omega \cong n_{2\omega} \cong 1.7$  and $\kappa_{2\omega} \cong 0.4$ for PDI8-CN$_2$  have been used. Considering a film thickness up to $5\cdot L_D$, maximum difference between the exact and approximated $\left|\chi _{zxx}(d)\right|$ value is found to be about 3\%, a result that can be considered satisfactory within the experimental uncertainties. We therefore used Eq.(6) to fit the experimental SHG susceptibilities $\left|\chi _{zxx}\right|(d)$ of the PDI8-CN$_2$ films. The results are reported in Fig.(6), where full circles and dashed line represent the experimental data (obtained by PR-SHG analysis, Eq.(1)) and the best-fit curve obtained by means of Eq.(6), respectively. As a reference, the average experimental value of $\left|\chi _{zxx}\right|$ amplitude of un-deposited substrates is also reported (empty square). The data fitting gives an estimated value of Debye length of $L_D$ = 9 nm, corresponding to about 4-5 monolayers (ML) of PDI8-CN$_2$ on average, with a experimental uncertainty of about 1 ML monolayer.

\begin{figure}
\includegraphics{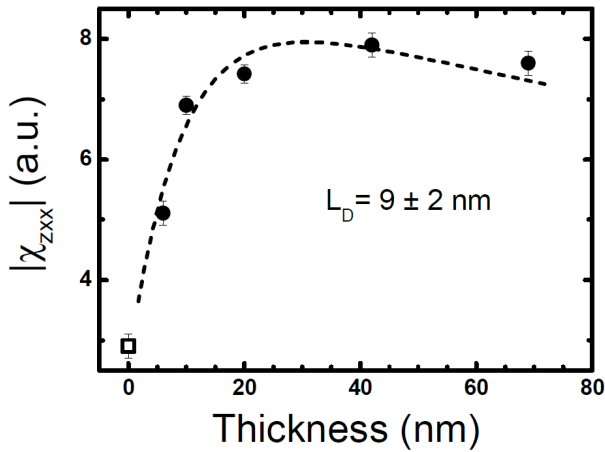}
\caption{(Color online) Full circles: $\left|\chi_{zxx}\right|$ SHG susceptibility obtained by PR-SHG analysis of the PDI8-CN$_2$ films. Dashed curve: best fit of experimental data obtained by Eq.(6). Empty square: average experimental value of un-deposited substrates $\left|\chi_{zxx}\right|$ amplitude.} 		
\end{figure}

\subsection{Additional observations and remarks on SiO$_2$/PDI8-CN$_2$ interfacial charge screening interpretation}

The model described in the previous sub-section assumes the presence of localized charges at SiO$_2$/PDI8-CN$_2$ interface and of mobile charges (able to spatially redistribute themselves) in the bulk of organic semiconductor. We discuss here additional observations supporting such assumptions.

Concerning the presence of mobile electrons in as-grown PDI8-CN$_2$ films, interesting considerations can be expounded from PL+PLE findings previously reported. Even if no specific references are available to elucidate the nature of states contributing to PL emission in PDI8-CN$_2$, it is worth citing a recent in-situ PL analysis performed on molecular films of perylene cores.\cite{Chen10} In this work, authors evidenced the different spectral contributions originating from excimers and from defects states at room temperature, with the former contributing to a PL peak significantly red-shifted ($\Delta E \approx$ 540 meV)  with respect to the 0-0 monomer transition.\cite{Chen10,Nishimura84} These results are very close to the ones here reported for PDI8-CN$_2$, whose PL also can be decomposed as a superposition of two contributes, one approximately peaked at $E_a$=1.76 eV (red dotted curve in Fig. 2) and another at about $E_b$=1.90 eV (blue dotted curve in Fig. 2). It is to be noted that the energy distance between $E_a$ and the first absorption transition ($E_1$) is $\left|E_1-E_a\right| \approx $ 520 meV, almost equal to the result reported by Chen and Richardson.\cite{Chen10} It is also to be underlined that the second emission band occurs at photon energies very close to the onset of optical absorption $E_0$ evidenced by the PLE spectra. On these basis, the PL components peaked at energies $E_a$ and $E_b$ can be reasonably assigned to excimer emission and below-bandgap defective states, similarly to Ref.49. This allows to sketch a scheme for PDI8-CN$_2$ states occupation at equilibrium in which defect states accumulating below the LUMO edge are prevalently occupied and absorption transitions at photon energy below the $E_0$ edge are prevented by state-filling (i.e. Pauli exclusion principle). In such a case, after optical excitation provided by photons with energy $\hbar\omega > E_0$, electrons occupying defective states below the LUMO edge and near band-edge states slightly above $E_0$ can recombine with holes created by photo-excitation, thus explaining why the near band-edge transitions at photon energy less than $E_0$ are observed in emission spectrum and not in excitation spectrum.

It is a key point that the above described occupation of near band-edge states implies a pinning of the Fermi level close to the LUMO edge and, as a consequence, a non-negligible occupation of LUMO states at finite temperature. This supports the presence of mobile electrons able to spatially redistribute themselves in as-grown PDI8-CN$_2$ films.

Other recent literature works also support this latter conclusion. A very direct evidence of LUMO occupation was obtained by recent ultraviolet photoemission spectroscopy investigations evidencing a Fermi level positioned very close to the LUMO edge in PDI8-CN$_2$ films.\cite{Aversa12} Furthermore, the presence of mobile bulk electrons in PDI8-CN$_2$ thin-film transistors using non-passivated SiO$_2$ layers as gate dielectric was also evidenced in Ref. 53 by observation of negative threshold voltages, accompanied by measurable source-drain currents occurring even for null gate voltage.\cite{DiGirolamo12b}

Finally, it is worth to briefly discuss about possible mechanisms leading to Debye-like charge redistribution here hypothesized, focusing in particular on specific interactions involving the organic semiconductor and the SiO$_2$ dielectric. In this regard, it is worth mentioning that the negative threshold voltage observed in Ref. 53 for PDI8-CN$_2$ OFETs using untreated SiO$_2$ as gate dielectric shifted toward less negative voltages after SiO$_2$ passivation by hexamethyldisilazane (HMDS). Due to the hydrophobic character of HMDS-passivated surfaces \cite{Mathijssen08}, authors suggested the occurrence in PDI8-CN$_2$ of a water-induced oxidation accompanied by proton formation, according to the redox reaction 2H$_{2}$O$+$4OS$\rightleftharpoons 4$H$^{+}+$4OS$^{-}+$O$_{2\text{(solv)}}$,  where OS and OS$^-$ indicate the organic semiconductor at neutral state and at negatively charged state. At equilibrium, a net amount of protons can diffuse in the SiO$_2$ layer barrier and be eventually back-diffused towards the semiconductor during application of positive gate bias, leading to source-drain current instability.\cite{DiGirolamo12b} The negatively-charged PDI8-CN$_2$ molecules (OS$^-$) can promote the electron transport, resulting in the source-drain currents observed even in absence of applied gate voltage. Such a proposed scheme would explain both the positive shift of threshold voltage and the reduction of mobile charge concentrations observed after HMDS treatment of SiO$_2$. It is worth noting that a similar mechanism has been proposed by Sharma and coworkers to explain the operational instabilities in p-type OFETs.\cite{Sharma10}
According to the above described interpretation, we hypothesize that the built-in electrostatic field contributing to SHG response can thus originate from OS$^-$ redistributing themselves to screen out the net charge associated to protons migrated in the SiO$_2$ layer. However, further investigations are required in order to confirm such interpretation.

\section{Conclusions}

In conclusion, in this work we investigated on the possibility to employ optical second-harmonic generation as a technique to perform non-destructive and spatially-selective probing of electronic response in PDI8-CN$_2$ molecular films deposited on SiO$_2$ dielectric layers. Evidences of nonlinear polarization contributions originating from charges lying at the buried PDI8-CN$_2$/SiO$_2$ interface and/or from internal built-in electrostatic fields formed in the organic semiconductor were sought by determination of nonlinear dielectric susceptibilities in PDI8-CN$_2$ films of variable thickness.
The experimental results pointed out a finite three-dimensional “SHG-active” semiconductor region in which spatial distribution of charge carriers lacked inversion symmetry, suggesting that purely-interfacial SHG did not represent the only contribution to the total SH intensity. In particular, we showed that the experimental findings could be modeled in terms of electric field-induced SHG, resulting from the presence of a net charge lying in the SiO$_2$ gate dielectric and inducing a non-uniform charge distribution in bulk PDI8-CN$_2$, according to a Debye-like charge screening mechanism. Application of our model for the electric field-dependent nonlinear susceptibility to the experimental data gave a screening length of about 4-5 ML characterizing the non-uniform charge distribution in PDI8-CN$_2$. Photoluminescence and excitation-resolved photoluminescence characterizations supported the occurrence of occupied LUMO states and the presence of mobile charge carriers in PDI8-CN$_2$, thus reinforcing the proposed interpretation. Reduction-oxidation reactions involving PDI8-CN$_2$ and water molecules were discussed as the possible source of the net charge inducing the screening field formation. The work outlines a useful and non-destructive method to probe charged layers at buried organic/dielectric interfaces.

\appendix
\section{Determination of total $\chi_{zxx}$ effective susceptibility}

In the present Appendix we determine the expression reported in Eq. (2) for the effective $\chi_{zxx}$ susceptibility in the presence of a spatially varying electrostatic $E^{dc}(z)$ field. The latter field gives rise to a contribution to the SH wave, whose expression can be obtained by means of the Green function formalism. As shown in Ref. 44, the output SH electric field $\textbf{E}_{2\omega }$ generated by a polarization $\mathbf{P}(z)$ and radiated in reflection geometry is given by:

\begin{equation}
	{{\bf{E}}_{2\omega }}(z) \propto \int_0^z {{\bf{P}}(z')\exp \left[ {i(z - z'){q_{2\omega }}} \right]dz'}]
	\label{eq:a1}
\end{equation}
We fix the origin of z axis ($z=0$) at SiO$_2$/PDI8-CN$_2$ buried interface, so that the amplitude of the output field is obtained by using $z = d$. Using the expression of electric field-induced polarization $\mathbf{P}_{2\omega }^{(EFISHG)}=\boldsymbol{\chi}^{(3)}:\mathbf{E}_{\omega}\mathbf{E}_{\omega}\mathbf{z}\cdot E^{dc}(z) $ (see also the main text) and Eq.(\ref{eq:a1}), one obtains:

\begin{equation}
\begin{split}
\mathbf{E}_{2\omega }^{(EFISHG)}&\propto \chi ^{(3)}:\mathbf{E}_{\omega}\mathbf{E}_{\omega }\mathbf{z}\times \\
	&\times\int_{0}^{d}\exp (iq_{2\omega }d)E^{dc}(z)\exp \left[ -i\left( q_{2\omega }+2q_{\omega }\right) z\right]dz \\
\end{split}
\label{eq:a2}
\end{equation}
where the fundamental and SH fields are expressed as $\mathbf{E}_{i}(x,y,z)=\mathbf{E}_{i}(x,y)\cdot\exp{(iq_{i}z)}$, so that $q_i$ = $\left({2\pi /{\lambda_i}}\right)\cdot {\tilde n_i}\cos{\theta _i}$ represent the wave-vector component perpendicular to sample surface of the i-th field ($i=\omega$ for the fundamental field, $i=2\omega$ for the SH field) and $\tilde n_i$ indicates the PDI8-CN$_2$ complex refractive index. It is to be noted that $\tilde{n}_{\omega}$ can be safely considered as a real quantity (as negligible optical absorption occurs in PDI8-CN$_2$ at the fundamental wavelength $\lambda_{\omega}$=1064 nm), while absorption is instead not negligible at SH wavelength $\lambda_{2\omega}$ = 532 nm and a finite imaginary part $\kappa_{2\omega}$ in the complex refractive index $\tilde{n}_{2\omega} = n_{2\omega }+ i\kappa_{2\omega}$ has to be considered.

Eq.(\ref{eq:a2}) can therefore be rewritten as:
\begin{equation}
\mathbf{E}_{2\omega}^{(EFISHG)} \propto {\boldsymbol{\chi}^{(3)}:\mathbf{E}_{\omega}\mathbf{E}_{\omega}\mathbf{z}\cdot \exp{\left(-d/\Lambda)\right)}F(d)}
\label{eq:a3}
\end{equation}
where
\[ F=\int_{0}^{d}E^{dc}(z)\exp{\left[ -iz/L_{C} \right]}dz \]
and where
\[ \Lambda=\lambda_{2\omega}/4\pi\left( n_{\omega}\cos_{\omega} + n_{2\omega}\cos_{2\omega}\right) \]
\[ L_{C}=\left|\operatorname{Re}(q_{2\omega })+ 2q_{\omega }\right| =\lambda _{\omega }/4\pi \left( n_{\omega }\cos \theta _{\omega }+n_{2\omega }\cos \theta _{2\omega } \right) \] are the optical attenuation length and the coherence length for reflected SH wave, respectively.

Eq.(\ref{eq:a3}) can be expressed in an equivalent manner by developing the tensor quantity on Cartesian components, as follows:

\begin{equation}
E_{i}^{(EFISHG)} \propto \left| {\chi _{ijkz}^{(3)}E_j^\omega E_k^\omega }\right| \exp(- d/\Lambda) \cdot F(d)
	\label{eq:a4}
\end{equation}
where the first member represents the i-th component of the electric field-induced SH wave and where the standard Einstein convention of summing over repeated indexes is used. A relation similar to (\ref{eq:a4}) can of course be given for the interface SH wave:

\begin{equation}
E_i^{(I)} \propto \left| \chi _{ijk}^{(I)}E_j^\omega E_k^\omega \right|
	\label{eq:a5}
\end{equation}
noting that in this latter case no integration over the z coordinate occurs, due to the local nature of the interfacial susceptibility. The total SH signal can thus be conveniently expressed as proportional to ${\left| {\chi _{ijk}^{(TOT)}E_j^\omega E_k^\omega } \right|^2}$, where $\chi _{ijk}^{(TOT)} = \chi _{ijk}^{(I)} + \chi _{ijkz}^{(3)}{e^{ - d/\Lambda }}F(d)$.

Being 3-rd rank tensors, both $\chi_{ijk}$ and the contracted $\chi_{ijkz}$ tensor (with forth index fixed to “z”) that appear in the previous expressions are described, in principle, by up to $3^3=27$ tensor elements. However, symmetry considerations can reduce significantly this number. In particular, in the case we are considering ($\infty mm$ symmetry) both tensors are described by only three non-null independent elements, namely $\chi_{zxx}$, $\chi_{xxz}$ and $\chi_{zzz}$ for the interfacial second-order susceptibility tensor and $\gamma_1=\chi_{zzxx}$, $\gamma_2=\chi_{zxzx}$ and $\gamma_3=\chi_{zxxz}$ for the $\chi^{(3)}$ tensor (the superscript “(3)” for the $\chi^{(3)}$ tensor elements has been omitted for the sake of brevity). From symmetry consideration the following equalities can be obtained:\cite{Butcher91}

\[\begin{array}{c}
\chi _{zxx}^{(I)} = \chi _{zyy}^{(I)}\\
\chi _{xxz}^{(I)} = \chi _{yyz}^{(I)} = \chi _{xzx}^{(I)} = \chi _{xzy}^{(I)}\\
{\gamma _1} = {\chi _{yyzz}} = {\chi _{zzyy}} = {\chi _{zzxx}} = {\chi _{xxyy}} = {\chi _{yyxx}}\\
{\gamma _2} = {\chi _{yzyz}} = {\chi _{zyzy}} = {\chi _{zxzx}} = {\chi _{xzxz}} = {\chi _{xyxy}} = {\chi _{yxyx}}\\
{\gamma _3} = {\chi _{yzzy}} = {\chi _{zyyz}} = {\chi _{zxxz}} = {\chi _{xzzx}} = {\chi _{xyyx}} = {\chi _{yxxy}}
\end{array}\]

The SH wave components can now be developed from Eq.(\ref{eq:a4}) and Eq.(\ref{eq:a5}) using the above listed non-null independent susceptibility elements. After calculation, the following expressions are obtained:

\begin{equation}
E_x^{(2\omega )} \propto 2E_x^\omega E_z^\omega  \cdot \left[ {\chi _{xxz}^{(I)} + \left( {\Gamma  - {\gamma _3}} \right){e^{ - d/\Lambda }}F(d)} \right]{\rm{ }}\\
\label{eq:a6}
\end{equation}

\begin{equation}
E_y^{(2\omega )} \propto 2E_y^\omega E_z^\omega  \cdot \left[ {\chi _{xxz}^{(I)} + \left( {\Gamma  - {\gamma _3}} \right){e^{ - d/\Lambda }}F(d)} \right]\\
\label{eq:a7}
\end{equation}

\begin{equation}
\begin{split}
E_z^{(2\omega )} \propto & \left[ {{{\left( {E_x^\omega } \right)}^2} + \left( {E_y^\omega } \right)} \right] \cdot \left[ {\chi _{zxx}^{(I)} + {\gamma _3}{e^{ - d/\Lambda }}F(d)} \right] + \\
& + {\left( {E_z^\omega } \right)^2} \cdot \left[ {\chi _{zzz}^{(I)} + \Gamma {e^{ - d/\Lambda }}F(d)} \right]
\end{split}
\label{eq:a8}
\end{equation}
where $\Gamma=\gamma_1+\gamma_2+\gamma_3$ and where the total SHG susceptibilities are the quantities in square parenthesis. In particular, from \ref{eq:a8} we obtain the equality reported in Eq.(\ref{eq:2}), i.e.:

	\[ \chi_{zxx}^{(Tot)}=\chi_{zxx}^{(I)}+\gamma_3\exp{\left(-d/\Lambda \right)}F\left(d\right).
\]

As a final check, it can be seen that in the s-in/p-out configuration ($\alpha=0$ in the $\alpha$p plots, corresponding to $E_x^\omega=E_z^\omega=0$) the equalities $E_x^{2\omega}=E_y^{2\omega}=0$ and $E_z^{2\omega} \propto \left(E_y^{2\omega}\right)^2 \cdot \chi_{zxx}$ are obtained by means of Eqs.(\ref{eq:a6}),(\ref{eq:a6}) and (\ref{eq:a8}). This implies that the p-polarized SH signal measured at $\alpha$=0 input polarization $I_p \propto \left| E_z^{2\omega} \right|^2$ is proportional to the square modulus of $\chi_{zxx}$, as also prescribed by Eq. (1).

\bibliographystyle{apsrev4-1}

\begin{thebibliography}{00}

\bibitem{Ishii99} H. Ishii, K. Sugiyama, E. Ito, and K. Seki, Advanced Materials 11, 605 (1999).
\bibitem{Braun09} S. Braun, W.R. Salaneck, and M. Fahlman, Advanced Materials 21, 1450 (2009).
\bibitem{Veres03} J. Veres, S. D. Ogier, S. W. Leeming, D. C. Cupertino, and S. Mohialdin Khaffaf, Advanced Functional Materials 13, 199 (2003).
\bibitem{Stassen04} A.F. Stassen, R.W.I. de Boer, N.N. Iosad, and A.F. Morpurgo, Applied Physics Letters 85, 3899 (2004).
\bibitem{Hulea06} I.N. Hulea, S. Fratini, H. Xie, C.L. Mulder, N.N. Iossad, G. Rastelli, S. Ciuchi, and A.F. Morpurgo, Nature Materials 5, 982 (2006).
\bibitem{Sirringhaus09} H. Sirringhaus, Advanced Materials 21, 3859 (2009).
\bibitem{Zilker01} S.J. Zilker, C. Detcheverry, E. Cantatore, and D.M. de Leeuw, Applied Physics Letters 79, 1124 (2001).
\bibitem{Richards08} T. Richards and H. Sirringhaus, Applied Physics Letters 92, 023512 (2008).
\bibitem{Sharma10} A. Sharma, S.G.J. Mathijssen, E.C.P. Smits, M. Kemerink, D.M. de Leeuw, and P.A. Bobbert, Phys. Rev. B 82, 075322 (2010).
\bibitem{Chen12}10 Y. Chen and V. Podzorov, Advanced Materials 24, 2679 (2012).
\bibitem{DeLeeuw97} D.M. de Leeuw, M.M.J. Simenon, A.R. Brown, and R.E.F. Einerhand, Synthetic Metals 87, 53 (1997).
\bibitem{Newman04} C.R. Newman, C.D. Frisbie, D.A. da Silva Filho, J.-L. Brédas, P.C. Ewbank, and K.R. Mann, Chem. Mater. 16, 4436 (2004).
\bibitem{Kobayashi04} S. Kobayashi, T. Nishikawa, T. Takenobu, S. Mori, T. Shimoda, T. Mitani, H. Shimotani, N. Yoshimoto, S. Ogawa, and Y. Iwasa, Nat Mater 3, 317 (2004).
\bibitem{Mathijssen07} S.G.J. Mathijssen, M. Cölle, H. Gomes, E.C.P. Smits, B. de Boer, I. McCulloch, P.A. Bobbert, and D.M. de Leeuw, Advanced Materials 19, 2785 (2007).
\bibitem{Mathijssen10} S.G.J. Mathijssen, M.J. Spijkman, A.M. Andringa, P.A. van Hal, I. McCulloch, M. Kemerink, R.A.J. Janssen, and D.M. de Leeuw, Advanced Materials 22, 5105 (2010).
\bibitem{Andringa12} A.M. Andringa, W.S. Christian Roelofs, M. Sommer, M. Thelakkat, M. Kemerink, and D.M. de Leeuw, Applied Physics Letters 101, 153302 (2012).
\bibitem{DiPietro12a} R. Di Pietro and H. Sirringhaus, Adv. Mater. 24, 3367 (2012).
\bibitem{DiPietro12b} R. Di Pietro, D. Fazzi, T.B. Kehoe, and H. Sirringhaus, J. Am. Chem. Soc. 134, 14877 (2012).
\bibitem{Butcher91} P.N. Butcher and D. Cotter, The Elements of Nonlinear Optics (Cambridge University Press, 1991).
\bibitem{Lettieri05} S. Lettieri, F. Gesuele, P. Maddalena, M. Liscidini, L.C. Andreani, C. Ricciardi, V. Ballarini, and F. Giorgis, Applied Physics Letters 87, 191110 (2005).
\bibitem{Lettieri07} S. Lettieri, F. Merola, P. Maddalena, C. Ricciardi, and F. Giorgis, Applied Physics Letters 90, 021919 (2007).
\bibitem{Chang97} Y.M. Chang, L. Xu, and H.W.K. Tom, Phys. Rev. Lett. 78, 4649 (1997).
\bibitem{Guo01} C. Guo, G. Rodriguez, and A.J. Taylor, Phys. Rev. Lett. 86, 1638 (2001).
\bibitem{Hoffmann11} T. Hoffmann, P. Thielen, P. Becker, L. Bohatý, and M. Fiebig, Phys. Rev. B 84, 184404 (2011).
\bibitem{Manaka07} T. Manaka, E. Lim, R. Tamura, and M. Iwamoto, Nature Photonics 1, 581 (2007).
\bibitem{Manaka08} T. Manaka, M. Nakao, E. Lim, and M. Iwamoto, Applied Physics Letters 92, 142106 (2008).
\bibitem{Lim08} E. Lim, H. Lee, T. Manaka, and M. Iwamoto, Japanese Journal of Applied Physics 47, 3179 (2008).
\bibitem{Tanaka11} Y. Tanaka, T. Manaka, and M. Iwamoto, Chemical Physics Letters 507, 195 (2011).
\bibitem{DiGirolamo12} F.V. Di Girolamo, M. Barra, F. Chiarella, S. Lettieri, M. Salluzzo, and A. Cassinese, Phys. Rev. B 85, 125310 (2012).
\bibitem{Jones04} B.A. Jones, M.J. Ahrens, M.H. Yoon, A. Facchetti, T.J. Marks, and M.R. Wasielewski, Angewandte Chemie International Edition 43, 6363 (2004).
\bibitem{Jones08} B.A. Jones, A. Facchetti, M.R. Wasielewski, and T.J. Marks, Advanced Functional Materials 18, 1329 (2008).
\bibitem{Yoo06} B. Yoo, A. Madgavkar, B.A. Jones, S. Nadkarni, A. Facchetti, K. Dimmler, M.R. Wasielewski, T.J. Marks, and A. Dodabalapur, IEEE Electron Device Letters 27, 737 (2006).
\bibitem{Rivnay09} J. Rivnay, L.H. Jimison, J.E. Northrup, M.F. Toney, R. Noriega, S. Lu, T.J. Marks, A. Facchetti, and A. Salleo, Nature Materials 8, 952 (2009).
\bibitem{Liscio12} F. Liscio, S. Milita, C. Albonetti, P. D’Angelo, A. Guagliardi, N. Masciocchi, R.G. Della Valle, E. Venuti, A. Brillante, and F. Biscarini, Advanced Functional Materials 22, 943 (2012).
\bibitem{Flytzanis95} Electronic delocalization over several identical repeat units is a fundamental prerequisite for large $\chi^{(3)}$ EFISHG susceptibilities. On this regard, see: C. Flytzanis ``Nonlinear Polarization'' in ``Optical and Non-linear Materials: Principles and Applications'', Proceedings of the International School of Physics ``Enrico Fermi'', IOS Press, The Netherlands (1995), ISBN: 9051992041.
\bibitem{Chiarella11} F. Chiarella, M. Barra, A. Cassinese, F. Di Girolamo, P. Maddalena, L. Santamaria, and S. Lettieri, Applied Physics A: Materials Science and Processing 104, 39 (2011).
\bibitem{Krug04} J. Krug, Physica A: Statistical Mechanics and Its Applications 340, 647 (2004).
\bibitem{Corn94} R.M. Corn and D.A. Higgins, Chemical Reviews 94, 107 (1994).
\bibitem{Guyot86} P. Guyot-Sionnest, W. Chen, and Y.R. Shen, Physical Review B 33, 8254 (1986).
\bibitem{Zhang90} T.G. Zhang, C.H. Zhang, and G.K. Wong, J. Opt. Soc. Am. B 7, 902 (1990).
\bibitem{Lee67} C.H. Lee, R.K. Chang, and N. Bloembergen, Physical Review Letters 18, 167 (1967).
\bibitem{Shen02} Y.R. Shen, The Principles of Nonlinear Optics, Reprint (Wiley-Interscience, 2002).
\bibitem{Aktsipetrov94} O.A. Aktsipetrov, A.A. Fedyanin, V.N. Golovkina, and T.V. Murzina, Opt. Lett. 19, 1450 (1994).
\bibitem{Sipe87} J.E. Sipe, J. Opt. Soc. Am. B 4, 481 (1987).
\bibitem{Tkachenko12} V. Tkachenko, F.V. Di Girolamo, F. Chiarella, A. Cassinese, and G. Abbate, Thin Solid Films 520, 2390 (2012).
\bibitem{Mayergoyz86} I.D. Mayergoyz, Journal of Applied Physics 59, 195 (1986).
\bibitem{Pacelli97} A. Pacelli, IEEE Transactions on Electron Devices 44, 1169 (1997).
\bibitem{Luth10} H. Luth, Solid Surfaces, Interfaces and Thin Films, 5th ed. (Springer, 2010).
\bibitem{Chen10} Q. Chen and N.V. Richardson, J. Phys. Chem. C 114, 6062 (2010).
\bibitem{NotaPL} Relaxed excimer states and defect states are separated by an energy barrier and thermal energy is therefore required to overcome it. As a consequence, an increase in temperature is accompanied by a decrease of excimer band and a relative increase of defect states contribution, as shown by Chen and coworkers (Ref. 49).
\bibitem{Nishimura84} H. Nishimura, T. Yamaoka, K. Mizuno, M. Iemura, and A. Matsui, Journal of the Physical Society of Japan 53, 3999 (1984).
\bibitem{Aversa12} L. Aversa, R. Verucchi, R. Tatti, F.V. Di Girolamo, M. Barra, F. Ciccullo, A. Cassinese, and S. Iannotta, Applied Physics Letters 101, 233504 (2012).
\bibitem{DiGirolamo12b} F.V. Di Girolamo, F. Ciccullo, M. Barra, A. Carella, and A. Cassinese, Organic Electronics 13, 2281 (2012).
\bibitem{Mathijssen08} S.G.J. Mathijssen, M. Kemerink, A. Sharma, M. Colle, P.A. Bobbert, R.A.J. Janssen, and D.M. de Leeuw, Advanced Materials 20, 975 (2008).

\end{thebibliography}

\end{document}